\documentclass{Pos}

\title{Nonstandard Higgs decays in the E$_6$SSM}

\ShortTitle{Nonstandard Higgs decays in the E6SSM}

\author{Jonathan Hall\\
        University of Southampton\\
        E-mail: \email{jonathan.hall@soton.ac.uk}}

\author{Steve F.~King\\
        University of Southampton\\
        E-mail: \email{s.f.king@soton.ac.uk}}

\author{\speaker{Roman Nevzorov}%
         \thanks{On leave of absence from the Theory Department, ITEP, Moscow, Russia.}\\
        University of Hawaii\\
        E-mail: \email{nevzorov@phys.hawaii.edu}}

\author{Sandip Pakvasa\\
        University of Hawaii\\
        E-mail: \email{pakvasa@phys.hawaii.edu}}

\author{Marc Sher\\
        College of William and Mary\\
        E-mail: \email{mtsher@wm.edu}}

\abstract{We study the nonstandard decays of the lightest Higgs state
within the Exceptional Supersymmetric Standard Model (E$_6$SSM).
We argued that the SM--like Higgs boson can decay predominantly into
dark matter particles while its branching ratios into SM particles
varies from 2\% to 4\%. This scenario also implies the presence
of other relatively light Inert chargino and neutralino states
in the particle spectrum with masses below $200\,\mbox{GeV}$.
We argue that in this case the decays of the lightest Higgs
boson into $l^{+} l^{-} + X$ may play an essential role in
the Higgs searches.}

\FullConference{The XIXth International Workshop on High Energy Physics
and Quantum Field Theory \\
                 8-15 September 2010\\
                 Golitsyno, Moscow, Russia}

\begin{document}

\section{E$_6$SSM}
The E$_6$SSM is based on the $SU(3)_C\times SU(2)_W\times U(1)_Y \times U(1)_N$
gauge group which is a subgroup of $E_6$ \cite{King:2005jy}-\cite{King:2005my}.
The additional low energy $U(1)_N$ is a linear superposition of $U(1)_{\chi}$
and $U(1)_{\psi}$, i.e. $U(1)_N=\frac{1}{4} U(1)_{\chi}+\frac{\sqrt{15}}{4} U(1)_{\psi}$.
Two anomaly--free $U(1)_{\psi}$ and $U(1)_{\chi}$ symmetries are defined by:
$E_6\to SO(10)\times U(1)_{\psi}\,,\quad SO(10)\to SU(5)\times U(1)_{\chi}$.
The extra $U(1)_N$ gauge symmetry is defined such that right--handed
neutrinos do not participate in the gauge interactions. Since right--handed
neutrinos have zero charges they can be superheavy, shedding light on the
origin of the mass hierarchy in the lepton sector and providing a mechanism
for the generation of the baryon asymmetry in the Universe via
leptogenesis \cite{King:2008qb}.

To ensure anomaly cancellation the particle content of the E$_6$SSM is
extended to include three complete fundamental $27$ representations of $E_6$
Each $27_i$ multiplet contains a SM family of quarks and leptons,
right--handed neutrino $N^c_i$, SM-type singlet fields $S_i$ which carry
non-zero $U(1)_{N}$ charge, a pair of $SU(2)_W$--doublets $H^d_{i}$ and
$H^u_{i}$ and a pair of colour triplets of exotic quarks $\overline{D}_i$ and
$D_i$ which can be either diquarks (Model I) or leptoquarks (Model II)
\cite{King:2005jy}-\cite{King:2005my}.\, $S_i$, $H^d_{i}$ and $H^u_{i}$ form
either Higgs or inert Higgs multiplets. In addition to the complete $27_i$
multiplets the low energy particle spectrum of the E$_6$SSM is supplemented
by $SU(2)_W$ doublet $H'$ and anti-doublet $\overline{H}'$ states from extra
$27'$ and $\overline{27'}$ to preserve gauge coupling unification.
The analysis performed in \cite{King:2007uj} shows
that the unification of gauge couplings in the E$_6$SSM can be achieved
for any phenomenologically acceptable value of $\alpha_3(M_Z)$ consistent
with the measured low energy central value. The presence of a $Z'$ boson
and of exotic quarks predicted by the E$_6$SSM provides spectacular new
physics signals at the LHC which were discussed in \cite{King:2005jy}--\cite{King:2005my},
\cite{King:2006vu}--\cite{King:2006rh}. Recently the particle
spectrum and collider signatures associated with it were studied within
the constrained version of the E$_6$SSM \cite{Athron:2008np}--\cite{Athron:2009bs}.

The superpotential in the $E_6$ inspired models involves many new Yukawa
couplings that induce non--diagonal flavour transitions. To suppress these effects
in the E$_6$SSM an approximate $Z^{H}_2$ symmetry is imposed. Under this symmetry
all superfields except one pair of $H^d_{i}$ and $H^u_{i}$ (say $H_d\equiv H^d_{3}$ and
$H_u\equiv H^u_{3}$) and one SM-type singlet field ($S\equiv S_3$) are odd. The $Z^{H}_2$
symmetry reduces the structure of the Yukawa interactions to
\begin{eqnarray}
W_{\rm E_6SSM}\simeq  \lambda \hat{S} (\hat{H}_u \hat{H}_d)+
\lambda_{\alpha\beta} \hat{S} (\hat{H}^d_{\alpha} \hat{H}^u_{\beta})+
\kappa_{ij} \hat{S} (\hat{D}_i\hat{\overline{D}}_j)+
f_{\alpha\beta} \hat{S}_{\alpha} (\hat{H}_d \hat{H}^u_{\beta})\nonumber\\[2mm]
+\tilde{f}_{\alpha\beta} \hat{S}_{\alpha} (\hat{H}^d_{\beta}\hat{H}_u)
+\mu'(\hat{H}'\hat{\overline{H'}})+h^{E}_{4j}(\hat{H}_d \hat{H}')\hat{e}^c_j
+W_{\rm MSSM} (\mu=0)\,,
\label{essm2}
\end{eqnarray}
where $\alpha,\beta=1,2$ and $i,j=1,2,3$\,. Here we assume that all right--handed
neutrinos are heavy. The $SU(2)_W$ doublets $\hat{H}_u$ and $\hat{H}_d$
and SM-type singlet field $\hat{S}$, that are even under the $Z^{H}_2$ symmetry,
play the role of Higgs fields. At the physical vacuum they develop VEVs
$\langle H_d\rangle = v_1/\sqrt{2}$, $\langle H_u\rangle =v_2/\sqrt{2}$
and $\langle S\rangle = s/\sqrt{2}$. Instead of $v_1$ and $v_2$ it is more
convenient to use $\tan\beta=v_2/v_1$ and $v=\sqrt{v_1^2+v_2^2}=246\,\mbox{GeV}$.
The VEV of the SM-type singlet field, $s$, breaks the extra $U(1)_N$ symmetry
thereby providing an effective $\mu$ term as well as the necessary exotic fermion
masses and also inducing that of the $Z'$ boson. In the E$_6$SSM the Higgs
spectrum contains one pseudoscalar, two charged and three CP--even states.
In the leading two--loop approximation the mass of the lightest CP--even
Higgs boson does not exceed $150-155\,\mbox{GeV}$ \cite{King:2005jy}.

\section{Lightest Inert neutralinos}
The neutral components of the Inert Higgsinos
($\tilde{H}^{d0}_1$, $\tilde{H}^{d0}_2$, $\tilde{H}^{u0}_1$, $\tilde{H}^{u0}_2$)
and Inert singlinos ($\tilde{S}_1$, $\tilde{S}_2$) mix and form Inert neutralino
states. In the field basis
$(\tilde{H}^{d0}_2,\,\tilde{H}^{u0}_2,\,\tilde{S}_2,\,\tilde{H}^{d0}_1,\,\tilde{H}^{u0}_1,\,\tilde{S}_1)$
the corresponding mass matrix takes a form
\begin{equation}
M_{IN}=
\left(
\begin{array}{cc}
A_{22}  & A_{21} \\[2mm]
A_{12}  & A_{11}
\end{array}
\right)\,,\qquad
A_{\alpha\beta}=-\frac{1}{\sqrt{2}}
\left(
\begin{array}{ccc}
0                                           & \lambda_{\alpha\beta} s           & \tilde{f}_{\beta\alpha} v \sin{\beta}
\\[2mm]
\lambda_{\beta\alpha} s                     & 0                                 & f_{\beta\alpha} v \cos{\beta} \\[2mm]
\tilde{f}_{\alpha\beta} v \sin{\beta}       & f_{\alpha\beta} v \cos{\beta}     & 0
\end{array}
\right)\,,
\label{icn1}
\end{equation}
so that $A_{12}=A^{T}_{21}$.
In the following analysis we shall choose the VEV of the SM singlet field to be large
enough ($s=2400\,\mbox{GeV}$) so that the experimental constraints on $Z'$ boson mass
($M_{Z'}\gtrsim 865\,\mbox{GeV}$) and $Z-Z'$ mixing are satisfied and all Inert chargino
states are heavier than $100\,\mbox{GeV}$. In addition, we also require the validity of
perturbation theory up to the GUT scale. The restrictions specified above
set very strong limits on the masses of the lightest Inert neutralinos.
In particular, our numerical analysis indicates that the lightest and second
lightest Inert neutralinos ($\chi^0_{1}$ and $\chi^0_{2}$) are typically
lighter than $60-65\,\mbox{GeV}$ \cite{10}--\cite{Hesselbach:2007te}. Therefore the
lightest Inert neutralino tends to be the lightest SUSY particle in the spectrum and
can play the role of dark matter. The neutralinos $\chi^0_{1}$ and $\chi^0_{2}$ are
predominantly Inert singlinos. Their couplings to the $Z$--boson can be rather small
so that such Inert neutralinos would remain undetected at LEP.

In order to clarify the results of our numerical analysis, it is useful to
consider a simple scenario when $\lambda_{\alpha\beta}=\lambda_{\alpha}\,\delta_{\alpha\beta}$,
$f_{\alpha\beta}=f_{\alpha}\,\delta_{\alpha\beta}$ and
$\tilde{f}_{\alpha\beta}=\tilde{f}_{\alpha}\,\delta_{\alpha\beta}$.
In the limit where off--diagonal Yukawa couplings vanish and
$\lambda_{\alpha} s\gg f_{\alpha} v,\, \tilde{f}_{\alpha} v$ the eigenvalues of the
Inert neutralino mass matrix can be easily calculated (see \cite{Hall:2009aj}).
In particular the masses of two lightest Inert neutralino states ($\chi^0_{1}$ and $\chi^0_{2}$)
are given by
\begin{equation}
m_{\chi^0_{\alpha}}\simeq \frac{\tilde{f}_{\alpha} f_{\alpha} v^2 \sin 2\beta}{2\, m_{\chi^{\pm}_{\alpha}}}\,.
\label{icn10}
\end{equation}
where $m_{\chi^{\pm}_{\alpha}}=\lambda_{\alpha} s/\sqrt{2}$ are masses of
the Inert charginos. From Eq.~(\ref{icn10}) one can see that the masses of $\chi^0_{1}$
and $\chi^0_{2}$ are determined by the values of the Yukawa couplings $\tilde{f}_{\alpha}$
and $f_{\alpha}$. They decrease with increasing $\tan\beta$ and chargino masses.
In this approximation the part of the Lagrangian, that describes interactions of Z with
$\chi^0_1$ and $\chi^0_2$, can be presented in the following form:
\begin{eqnarray}
\mathcal{L}_{Z\chi\chi}=\sum_{\alpha,\beta}\frac{M_Z}{2 v}
Z_{\mu}\biggl(\chi^{0T}_{\alpha}\gamma_{\mu}\gamma_{5}\chi^0_{\beta}\biggr)
R_{Z\alpha\beta}\,,\qquad\qquad\qquad\\[0mm]
R_{Z\alpha\beta}=R_{Z\alpha\alpha}\,\delta_{\alpha\beta}\,,\qquad
R_{Z\alpha\alpha}=\frac{v^2}{2 m_{\chi^{\pm}_{\alpha}}^2}
\biggl(f_{\alpha}^2\cos^2\beta-\tilde{f}_{\alpha}^2\sin^2\beta\biggr)\,.
\label{icn14}
\end{eqnarray}
Eqs.~(\ref{icn14}) demonstrates that the couplings of $\chi^0_1$ and $\chi^0_2$ to
the Z-boson can be very strongly suppressed or even tend to zero. This happens when
$|f_{\alpha}|\cos\beta\approx |\tilde{f}_{\alpha}|\sin\beta$.

Although $\chi^0_1$ and $\chi^0_2$ might have extremely small couplings to
$Z$, their couplings to the lightest CP--even Higgs boson $h_1$ can not be
negligibly small if the corresponding states have appreciable masses. When the SUSY
breaking scale $M_S$ and the VEV $s$ of the singlet field are considerably larger
than the EW scale, the mass matrix of the CP--even Higgs sector has a hierarchical
structure and can be diagonalised using the perturbation theory
\cite{Nevzorov:2001um}--\cite{Nevzorov:2000uv}. In this case the
lightest CP--even Higgs state is the analogue of the SM Higgs field and is responsible
for all light fermion masses in the E$_6$SSM. Therefore it is not so surprising that
in the limit when $\lambda_{\alpha} s\gg f_{\alpha} v,\, \tilde{f}_{\alpha} v$
the part of the Lagrangian that describes the interactions of $\chi^0_1$ and $\chi^0_2$
with $h_1$ takes a form
\begin{equation}
\mathcal{L}_{H\chi\chi}=\sum_{\alpha,\beta} (-1)^{\theta_{\alpha}+\theta_{\beta}}
X^{h_1}_{\alpha\beta} \biggl(\psi^{0T}_{\alpha}(-i\gamma_{5})^{\theta_{\alpha}+
\theta_{\beta}}\psi^0_{\beta}\biggr) h_1\,,\qquad
X^{h_1}_{\gamma\sigma}\simeq\frac{|m_{\chi^0_{\sigma}}|}{v}\,\delta_{\gamma\sigma}\,,
\label{icn12}
\end{equation}
i.e. the couplings of $h_1$ to $\chi^0_1$ and $\chi^0_2$ is proportional to the mass/VEV.
In Eq.~(\ref{icn12}) $\psi^0_{\alpha}=(-i\gamma_5)^{\theta_{\alpha}}\chi^0_{\alpha}$ is
the set of Inert neutralino eigenstates with positive eigenvalues, while $\theta_{\alpha}$
equals 0 (1) if the eigenvalue corresponding to $\chi^0_{\alpha}$ is positive (negative).

\begin{table}
\centering
\begin{tabular}{|c|c|c|c|c|}
\hline
                             & A      &  B       &  C       & D\\
\hline
$\lambda_{22}(M_t)$          & 0.094  & 0.001    & 0.001    & 0.468\\
\hline
$\lambda_{21}(M_t)$          & 0      & 0.094    & 0.079    & 0.05\\
\hline
$\lambda_{12}(M_t)$          & 0      & 0.095    & 0.08     & 0.05\\
\hline
$\lambda_{11}(M_t)$          & 0.059  & 0.001    & 0.001    & 0.08\\
\hline
$f_{22}(M_t)$                & 0.53   & 0.001    & 0.04     & 0.05\\
\hline
$f_{21}(M_t)$                & 0.053  & 0.69     & 0.68     & 0.9\\
\hline
$f_{12}(M_t)$                & 0.053  & 0.69     & 0.68     & 0.002\\
\hline
$f_{11}(M_t)$                & 0.53   & 0.001    & 0.04     & 0.002\\
\hline
$\tilde{f}_{22}(M_t)$        & 0.53   & 0.001    & 0.04     & 0.002\\
\hline
$\tilde{f}_{21}(M_t)$        & 0.053  & 0.49     & 0.49     & 0.002\\
\hline
$\tilde{f}_{12}(M_t)$        & 0.053  & 0.49     & 0.49     & 0.05\\
\hline
$\tilde{f}_{11}(M_t)$        & 0.53   & 0.001    & 0.04     & 0.65\\
\hline
$|m_{\chi_1^0}|$/GeV         & 35.42  & 45.60    & 45.1     & 46.24\\
\hline
$|m_{\chi_2^0}|$/GeV         & 51.77  & 45.95    & 55.3     & 46.60\\
\hline
$|m_{\chi_3^0}|$/GeV         & 105.3  & 158.7    & 133.3    & 171.1\\
\hline
$|m_{\chi_4^0}|$/GeV         & 152.7  & 162.1    & 136.9    & 171.4\\
\hline
$|m_{\chi_5^0}|$/GeV         & 162.0  & 204.7    & 178.4    & 805.4\\
\hline
$|m_{\chi_6^0}|$/GeV         & 201.7  & 207.8    & 192.2    & 805.4\\
\hline
$|m_{\chi_1^{\pm}}|$/GeV     & 100.1  & 158.5    & 133      & 125.0\\
\hline
$|m_{\chi_2^{\pm}}|$/GeV     & 159.5  & 162.3    & 136.8    & 805.0\\
\hline
$\Omega_{\chi} h^2$          & 0.107  & 0.0886   & 0.0324   & 0.00005\\
\hline
$R_{Z\chi_1^0\chi_1^0}$      & -0.115 & -0.0205  & -0.0217  & -0.0224\\
\hline
$R_{Z\chi_2^0\chi_2^0}$      & -0.288 & -0.0208  & -0.0524  & -0.0226\\
\hline
$R_{Z\chi_1^0\chi_2^0}$      &  0.046 &  0.0015  & -0.0020  & -0.213\\
\hline
$Br(h\to \chi_2\chi_2)$      & 20.3\% & 48.0\%   & 12.3\%   & 47.9\%\\
\hline
$Br(h\to \chi_2\chi_1)$      & 0.25\% & 0        & 0        & 0\\
\hline
$Br(h\to \chi_1\chi_1)$      & 76.3\% & 49.5\%   & 83.4\%   & 49.3\%\\
\hline
$Br(h\to b\bar{b})$          & 2.82\% & 2.28\%   & 3.93\%   & 2.56\%\\
\hline
$Br(h\to \tau\bar{\tau})$    & 0.30\% & 0.24\%   & 0.42\%   & 0.27\%\\
\hline
$\Gamma^{tot}$/MeV           & 81.7   & 101      & 58.6     & 89.8\\
\hline
\end{tabular}
\caption{Benchmark scenarios. The branching ratios, masses of the
Inert neutralinos and charginos and couplings of the Inert neutralinos
are calculated for $s=2400\,\mbox{GeV}$, $\tan\beta=1.5$, $\lambda=g'_1=0.468$,
$A_{\lambda}=600\,\mbox{GeV}$, $m_Q=m_U=M_S=700\,\mbox{GeV}$, $X_t=\sqrt{6} M_S$ that
correspond to the two--loop lightest Higgs boson mass $m_{h_1}\simeq 116\,\mbox{GeV}$.
The one--loop masses of the heavy Higgs states are $m_A\simeq m_{H^{\pm}}\simeq m_{h_3}=1145\,\mbox{GeV}$
and  $m_{h_2}\simeq M_{Z'}\simeq 890\,\mbox{GeV}$.}
\end{table}

\section{Exotic Higgs decays}
In our analysis we required that the lightest Inert neutralino account for all or some of
the observed dark matter relic density. This sets another stringent constraint on the masses
and couplings of $\chi_1^0$. Indeed, because the lightest Inert neutralino states are almost
Inert singlinos, their couplings to the gauge bosons, Higgs states, quarks (squarks)
and leptons (sleptons) are rather small resulting in a relatively small annihilation
cross section of $\tilde{\chi}^0_1\tilde{\chi}^0_1\to \mbox{SM particles}$ and the possibility
of an unacceptably large dark matter density. A reasonable density of dark matter can be
obtained only for $|m_{\chi^0_{1}}|\sim M_Z/2$ when the lightest Inert neutralino states
annihilate mainly through an $s$--channel Z-boson, via its Inert Higgsino doublet components
which couple to the $Z$--boson. If $\tilde{\chi}^0_1$ annihilation proceeds through the
$Z$--boson resonance, i.e. $2|m_{\chi^0_{1}}|\approx M_Z$, then an appropriate value of
dark matter density can be achieved even for a relatively small coupling of $\tilde{\chi}^0_1$
to $Z$. Since the masses of $\chi^0_1$ and $\chi^0_2$ are much larger than the $b$--quark mass
and the decays of $h_1$ into these neutralinos are kinematically allowed, the SM--like Higgs
boson decays predominantly into the lightest inert neutralino states and has very small branching
ratios ($2\%-4\%$) for decays into SM particles.

In order to illustrate the features of the E$_6$SSM mentioned above we specify a set of
benchmark points (see Table 1). Within the E$_6$SSM relatively large masses for the lightest
and the second lightest Inert neutralinos can be obtained only for moderate values of
$\tan\beta\lesssim 2$. Even for $\tan\beta\lesssim 2$ the lightest Inert neutralino states
can get masses $\sim M_Z/2$ only if at least one light Inert chargino state and two Inert
neutralinos states, which are predominantly components of the $SU(2)_W$ doublet, have masses
below $200\,\mbox{GeV}$ (see Table 1). The Inert chargino and neutralinos states that are
mainly Inert Higgsinos couple rather strongly to $W$ and $Z$--bosons and therefore can be
efficiently produced at the LHC and then decay into the LSP and pairs of leptons and quarks
giving rise to remarkable signatures which can be observed in the near future.

When $\tan\beta\lesssim 2$ the mass of the lightest CP--even Higgs boson is very sensitive
to the choice of the coupling $\lambda(M_t)$. In particular, to satisfy LEP constraints
$\lambda(M_t)$ must be larger than $g'_1\simeq 0.47$. Large values of $\lambda(M_t)$ lead
to a qualitative pattern of the Higgs spectrum which is rather similar to the one which
arises in the PQ symmetric NMSSM \cite{Miller:2005qua}--\cite{Nevzorov:2004ge}.
In the considered limit the heaviest CP--even, CP--odd and charged states are almost
degenerate around $m_A$ and lie beyond the $\mbox{TeV}$ range while the mass of the
second lightest CP--even Higgs state is set by $M_{Z'}$ \cite{King:2005jy}.

If the masses of $\chi^0_1$ and $\chi^0_2$ are very close then the decays of
$h_1$ into $\chi_{\alpha}\chi_{\beta}$ will not be observed at the LHC giving rise to
a large invisible branching ratio of the SM--like Higgs boson. When the mass difference
between the second lightest and the lightest Inert neutralinos is larger than
$10\,\mbox{GeV}$ the invisible branching ratio remains dominant but some of the
decay products of $\chi_2$ might be observed at the LHC. In particular, there is a
chance that soft $\mu^{+} \mu^{-}$ pairs could be detected. Since the branching ratios
of $h_1$ into SM particles are extremely suppressed, the decays of the SM--like
Higgs boson into $l^{+} l^{-} + X$ could be important for Higgs searches \cite{10}.

\end{document}